\documentclass[12pt]{article}

\usepackage{makeidx}

\font\dun=cmdunh10 scaled 3583 

\font\dn=cmdunh10 scaled \magstep3
\font\dunh=cmdunh10 scaled \magstep1

\font\mit=cmti10 scaled \magstep1
\font\rit=cmti10 scaled \magstep3
\font\rrm=cmr10 scaled \magstep3

\font\manual=manfnt scaled \magstep1

\def\EinS{{\dunh EinS}}

\def\Mathematica{{\it Mathematica}}
\def\dbend{{\manual\char127}}

\def\expe{\dbend}
\def\begi{{$\clubsuit$}}

\def\bc{\begin{center}}
\def\ec{\end{center}}
\def\be{\begin{enumerate}}
\def\ee{\end{enumerate}}
\def\bi{\begin{itemize}}
\def\ei{\end{itemize}}

\def\bbbs{\bigskip\par\bigskip\par\bigskip}
\def\bbs{\bigskip\par\bigskip}
\def\bs{\bigskip}
\def\ms{\medskip}
\def\ssk{\smallskip}

\makeindex

\begin{document}

\tolerance=5000

\pagestyle{empty}

\vbox to 3cm{}

\bc

\dun EinS:\\
\bs
\rrm a \rit Mathematica \rrm package\\
\ms
\rrm for computations with indexed objects

\ec

\bbs

\bc

Version 2.6

\ec

\bbbs
\bbbs
\bbs

\bc
\dn Sergei A. Klioner
\ec

\bbs

\bc

\mit
Institute of Applied Astronomy, Russian Academy of Sciences,\\
8, Zhdanovskaya St., 197042 St.Petersburg, Russia\\
\ssk
Internet: klioner@ipa.rssi.ru

\ms

\rm and\\

\ms

\mit
Lohrmann Observatorium, Technische Universit\"at Dresden,\\
Mommsenstr., 13, 01062 Dresden, Germany\\
\ssk
Internet: klioner@rcs.urz.tu-dresden.de

\ec

\bbbs
\bbbs
\bbbs

\bc
9 June 1997
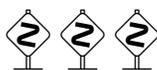
\ec

\bbs

\bc
\dbend\ \dbend\ \dbend
\ec

\newpage

\pagestyle{plain}

\label{license}

\vbox to 3cm{}

\bc

LICENSE AGREEMENT:

\ec

You are free to use \EinS\  for non-commercial use IF:

\ssk

\hskip 0.5cm NO FEE IS CHARGED FOR USE, COPYING OR DISTRIBUTION.

\ssk

\hskip 0.5cm IT IS NOT MODIFIED IN ANY WAY.

\bbs

The author disclaims all warranties as to this software, whether
express or implied, including without limitation any implied warranties
of merchantability, fitness for a particular purpose, functionality or
data integrity or protection.

\bbs

The inclusion of \EinS\  as part of a software and/or hardware package to
be distributed as well as the use of \EinS\  in a commercial environment
always require a special agreement. Please, contact the author for more
information.

\bbbs
\bbbs

If \EinS\  helped you to get results which you are going to publish,
a reference to an original description of the package is MANDATORY:

\ms

Klioner, S.A. (1995): \EinS: a \Mathematica\ package for tensorial
calculations in astronomical applications of relativistic gravity
theories. In Francaviglia, M. (ed.): {\it Abstracts of GR14,
the 14th international conference on general relativity and gravitation},
p. A.182. SIGRAV-GR14, Turin

\newpage

\bbbs
\bbbs
\bbbs

\bc
\bf Abstract
\ec

This report contains a description of \EinS\  version 2.6, which is
a \Mathematica\ package for computations with indexed objects.

\newpage

\tableofcontents

\newpage

\section{General description}

\EinS\  is a software allowing to perform operation on indexed objects,
which may or may not be tensors. \EinS\  is a package for \Mathematica,
which is one of the most advanced computer algebra systems. \EinS\  is a
relatively small package consisting in approximately 3000 lines which
constitute about 70 procedures. It was our intention to keep the
package [relatively] small. The general design and functionality of
\EinS\  resulted from scientific problems in the field of astronomical
applications of metric gravity theories tackled by the author during
last several years. Although, the first version of \EinS\  appeared as
early as in 1994 and one can say that \EinS\  is being developed
more than 2 years, the development of \EinS\  was not a significant
separate area of our scientific activity.  That is why, potential user
should not expect that \EinS\  is too general and flexible. For example,
we did not try to implement built-in procedures for computing all
standard quantities in General Relativity (curvature tensor, etc. (see,
however, Section \ref{future} below)).  Moreover, current version of
\EinS\  does not distinguish automatically even covariant and
contravariant indices. The main application field of \EinS\  is
computations with indexed objects involving implicit (Einstein)
summations (\EinS\  stands for "Einstein Summation handler").  \EinS\  is
designed to handle implicit summation rule, automatic generating of
dummy indices and has an efficient algorithm for simplification of
expressions containing dummy indices and objects with user-defined
symmetries.  The idea of the package was to create a simple, flexible
package which would be easy to alter for solving any problem involving
indexed objects and which would not waste time for doing anything we do
not ask it to do (the latter is usual fate of the users of too general
software). On the other hand, package works under \Mathematica\ which is
one of the most advanced and flexible computer algebra systems. This
allows the user to employ all the power of \Mathematica\ for solving his
problem. In order to work with \EinS\  user should have basic knowledge of
\Mathematica's control structures and language.

\EinS\  is not intended to perform very long computations since as well
as \Mathematica\ itself. The most laborious computations which we have
done with the use of \EinS\  was computing Landau-Lifshits pseudotensor
for a general metric containing terms of order of ${\cal O}(c^{-1})$ in
$g_{0i}$. The result contained about 2000 terms in total. Each term was
a product of up to 10 indexed object, some of which were defined to
symmetric. This calculation took about 6 hours on IBM PC 486DX33 with
16 Mbytes RAM. Recent progress in hardware probably allows to operate
with about 5000 or more terms depending on their structure within
\EinS, RAM being principal limitation.  However, if you have to operate
with longer expressions you have to turn to STENSOR (see, e.g.,
\cite{hart:96}).

For the moment there are only two [semi-]official publications
concerning \EinS\  \cite{klio:95a,klio:95}. If \EinS\  helped you to derive
a publishable results please refer to the official description of the
package as described on page \pageref{license}. We would also appreciate
receiving a reference on such publications.

Any reports of bugs would be appreciated and I will do my best
in fixing them.

\newpage

\section{Why one more package for indicial tensor calculations?}

It is well known that there exists a dozen of software systems and
packages for doing tensorial computations on a computer (see,
\cite{mac:87,hart:96} for comprehensive reviews). It is necessary to
distinguish between indicial tensorial computation and computation of
components of tensors. These two class of packages are quite different.
The latter usually allow one to compute a standard set of  quantities
for given components of metric: components of Christoffel symbols,
components of curvature tensor etc. The examples of such packages are
SHEEP (including CLASSI), EXCALC for REDUCE, CTENSR for Macsyma, and
recently developed GRTensor/II for MAPLE, and TTC and CARTAN for
\Mathematica.  On the other hand, the principal aim of a package for
indicial tensorial computations is to handle implicit summation and to
be able to simplify expressions involving dummy indices and
user-defined symmetries of various objects. Among them one could
mention the packages appeared in late the 1970s: ITENSR for Macsyma and
STENSOR for SHEEP and two newer packages for \Mathematica:  MathTensor
and RICCI.  SHEEP is a special-purpose system, and cannot handle all
variety of operations which may be necessary for the user.  REDUCE and
Macsyma are
relatively old system and are becoming probably obsolete (although the
latest version of Macsyma 2.2 for MS Windows looks promising its
performance cannot compete with that of Maple and even with
\Mathematica).  RICCI is designed for applications in differential
geometry. Although it is sometimes possible to force it to do the
computation similar to those described above, its performance
(especially its simplification algorithm) is too slow to handle
expressions containing many thousands of terms as we have in our
application field. MathTensor is a very large package requiring a lot
of additional hardware resources. This makes it difficult for us to use
MathTensor for deriving expressions which themselves take sometimes 10
Mbytes of RAM within \Mathematica.  Besides that neither of the
packages provide us with 3+1 split of indices which is so usual and
convenient in our application field restricted mainly by post-Newtonian
approximations of metric gravity theories.  All these arguments forced
us to implement our own small and flexible package to work under
\Mathematica\ which is one of the best modern computer algebra systems.
There are not so many programs (5--7 in total)which allow one to
operate with indicial tensor expressions, and every new such program is
of interest. This is the reason why even a small summary published
about \EinS\  attracted certain attention and a reference on \EinS\
appeared in a recent review on applications of computer algebra in
general relativity \cite{hart:96}.

\newpage

\section{Commands}

Below we describe all the commands of \EinS\  grouped by their
functionality. It is a good idea to look first at the demo programs
shipped with \EinS\  in order to understand better the logic, purposes
and functionality of the package.  \EinS\  have online help messages
accessible during \Mathematica\ session as usual. For example, the
command {\tt ?DefObject} prints a help message concerning the procedure
{\tt DefObject}. The commands of \EinS\  can
be divided into the following three groups:

\bi

\item
{\bf General purpose commands} which are not directly related to the purpose
of the package and extends \Mathematica\ functionality. We gathered these
functions in Section \ref{generalpurpose}.

\item
{\bf Commands for beginners} represent a minimal set of command
sufficient to perform any calculations which \EinS\  is originally
intended for (but sometimes not in the most effective way). These
commands are distributed over Sections \ref{input}--\ref{debug}
below and tagged with the sign \begi.

\item
{\bf Commands for experienced users} are commands which could be used
by an experienced user who wants to improve the performance of his
programs as well as to push \EinS\  to perform some calculation
for which \EinS\  is not intended originally. These
commands are distributed over the sections below and tagged with
the sign

$$\vbox{\hbox{\expe}\vskip 11pt}$$

\noindent
which "warns of a dangerous bend in the train of thought".
Following Donald Knuth's \TeX{}book we suggest not to use this
commands unless you really need.

\item
We describe also some of the {\bf internal commands}. A description of
these commands gives user some insight to the internal representation of
the data within \EinS\  as well as allows one to write own procedures
extending functionality of the package. These commands are described
in Section \ref{internal}.

\ei

\subsection{Getting started}
\label{load}
\index{Loading \EinS}
\index{Mathematica 3.0}

\EinS\  is shipped in form of encoded \Mathematica\ package.
You can directly read it into \Mathematica\ by the command

$$
\hbox{\tt <<EinS.m}
$$

Be sure that the directory where you placed file {\tt EinS.m} is in the
search path governed by variable {\tt \$Path} within \Mathematica.
After that the \EinS\  logo and its version appear as well as information
on \Mathematica's version on which it is being run.  If version of
Mathematica on which \EinS\  is being run does not coincide with that on
which we have tested \EinS\  (currently 2.1, 2.2 and 3.0),
a warning message is printed. You can now
proceed with typing commands of either \Mathematica\ itself or \EinS.
You cannot load \EinS\  twice within one \Mathematica\ session. If you
really need for some reasons (the author would appreciate hearing about
such reasons) to load \EinS\ again, you should restart \Mathematica's
kernel.

\subsection{General purpose commands}
\label{generalpurpose}

\EinS\  extends \Mathematica\ with a few general purpose functions which
are not related directly to the calculation which \EinS\  performs.  Some
of them are used internally by \EinS, but nevertheless can be probably
useful for users.

\bi

\ms\item[]{\tt SeriesLength}
\index{SeriesLength}

{\tt SeriesLength[expression]} prints number of terms in the expression
interpreting a product as one term (that is, number of monomials in the
expression). It also prints the number of terms in a series separately
at each power of the small parameter.

\index{Series within \EinS}

\ms\item[]{\tt SaveExpression}
\index{SaveExpression}

{\tt SaveExpression[file\_name,name]} saves the unevaluated name, equality
sign and the value into the specified file.
{\tt SaveExpression} appends its output to the specified file.
It does not use the Mathematica's
functions {\tt Definition} or {\tt FullDefinition}. It works faster
than SaveDefinition with the same arguments, but doesn't save the whole
Definition of a symbol. This command is especially suitable when you
work almost out of your RAM and want to save your result. Usual
\Mathematica's function {\tt Save} (which saves {\tt FullDefinition} of
the specified object) invoke a search throughout all the \Mathematica's
memory which may result in painful swapping.

See also: {\tt SaveDefinition}

\ms\item[]{\tt SaveDefinition}
\index{SaveDefinition}

{\tt SaveDefinition[file\_name,name]} saves Definition of name into the
specified file. {\tt SaveDefinition} appends its output to the
specified file. If you want to save the value of specific expression,
it is faster to use SaveExpression with the same arguments.  Note that
the \Mathematica's function {\tt Save} saves {\tt FullDefinition} of
the specified object.

See also: {\tt SaveExpression}

\ms\item[]{\tt KSubsets}
\index{KSubsets}

{\tt KSubsets[l,k]} returns all subsets of set l containing exactly k
elements, ordered lexicographically.

This function has been borrowed from the package {\tt
DiscreteMath`Combinatorica`} written by Steven S. Skiena to make \EinS\
independent of a larger package. The code is gratefully acknowledged.

\ms\item[]{\tt ExpandSeries}
\index{ExpandSeries}

{\tt ExpandSeries[expression]}
simplifies when possible nested calls to {\tt SeriesData}
which represent power series within \Mathematica.
{\tt ExpandSeries} allows to substitute a power series for a variables which
itself appear within a power series. For example,

\index{Series within \EinS}

\ssk

\begin{verbatim}
ExpandSeries[Series[1/(1+c x),{c,0,2}] /.
                           x -> Series[1/(1-c),{c,0,5}]]
\end{verbatim}

\ssk

gives correct result

\ssk

\begin{verbatim}
            3
1 - c + O[c].
\end{verbatim}

\ei

\subsection{How to input objects and formulas}
\label{input}

In order to do any calculations it is necessary to input
the object with which you want to work, to describe their
properties and to input their definition. \EinS\  has a number
of functions allowing to do so.

\bi

\ms\item[\begi]{\tt \$ESDimension}
\index{\$ESDimension}

{\tt \$ESDimension} is an integer positive constant defining dimension of space,
in which the user wants to work. It influences simplification of
the trace of the Kronecker symbol as well as the function ToComponents.

Default: {\tt \$ESDimension = 4}

\ms\item[\begi]{\tt DefObject}
\index{DefObject}

\begin{verbatim}
DefObject[object_name,valence,"print_name",
          {i1,i2,...},{j1,j2,...},
          {l1,...},ESIndicesLow -> True]
\end{verbatim}

defines the object of specified valence to be symmetric with respect to
indices number {\tt \{i1, i2, ...\}} and alternating with respect to
indices number {\tt \{j1, j2, ...\}}. Both list are optional and empty by
default.

The procedure also defines output formats for the object.  The 3rd
argument defines "print name" of the object with the specified valence.
This should be a \Mathematica's string. It is used for generating
output format used by {\tt PrintES} and any other printing utilities of
\Mathematica, as well as TeX format which is used by {\tt EinS2TeX} and
{\tt TeXSave}. The argument is optional. By default, name of the object
itself is used as its print name. User can define another print name for
\TeX\ output (for example, it is usually desirable to print
objects whose name is the name of a greek character with their \TeX\ names
({\tt \char'134{}omega}\ \/  for object {\tt omega}). It can be done manually by the user
by giving command

\begin{verbatim}
EinS2TeXString[omega]="\omega"
\end{verbatim}

However, this manually defined \TeX\ name shadows all other print names
for objects with given name irrespective of valence until it is removed
manually. One can do this with the command

\begin{verbatim}
EinS2TeXString[omega]=.
\end{verbatim}

The last two arguments of {\tt DefObject}
regulate printing of indices as superscripts or
subscripts.  The list \{l1,l2,...\} (empty by default) specifies
numbers of indices which will be printed by commands {\tt PrintES} (and
other \Mathematica's printing utilities) and {\tt EinS2TeX} as
subscripts, while all other indices will be printed as superscripts.
The 6th parameter which is also optional could be the rule {\tt
ESIndicesLow -> True}. This makes all subscripts to be superscripts and
vice verse.

Note that objects with the same input name, but with different valences
can have different symmetry properties, print names, and output and
\TeX\ formats. However, we do not recommend to use this opportunity
unless you are an experienced \Mathematica\ user, since defining several
objects with the same input name and different valences could sometimes
require from the user to take extra care in choosing order in which
various definitions are defined. This is a drawback of \EinS, but
rather a consequence of the general structure of \Mathematica\ language.
On the contrary, defining several objects with different input names
and valences, but the same print name is absolutely "safe".

{\tt DefObject} prints detailed report of the definition of object
resulted from current call.

\ms\item[\begi]{\tt ESIndicesLow}
\index{ESIndicesLow}

{\tt ESIndicesLow} is a parameter for the function {\tt DefObject}
controlling if indices of the object being defined will be printed as
subscripts or superscripts.  If {\tt ESIndicesLow -> False} (default)
all the indices which numbers are not specified in the 5th parameter of
{\tt DefObject} are printed as superscripts, while all other indices
are printed as subscripts.  If {\tt ESIndicesLow -> True} indices are
printed vice verse.

See also: {\tt DefObject}, {\tt PrintES}, {\tt SaveTeX}

\ms\item[\begi]{\tt DefES}
\index{DefES}

{\tt DefES[expression,
{list\_of\_variables\_to\_be\_interpreted\_as\_dummy\_indices}]}
\hfil\break
returns \EinS's internal representation of the expression containing
Einstein (implicit) sums. The second arguments represent a list of
names of variables which are to be interpreted as dummy indexes in the
specified expressions. You should take care to define these names to be
local names (for example, by calling {\tt DefES} from within
\Mathematica's {\tt Module} where names of dummy indices are defined to
be local. This is especially recommended if you make delayed
definitions via {\tt :=}. Note that after evaluation of {\tt DefES}
each dummy index received it own unique name and all the information on
the name with which you named it in the input expression of DefES is
lost.

See also:  {\tt ESRange}, {\tt \$ESDebug}, {\tt CheckDummies}

\ms\item[\begi]{\tt ESRange}
\index{ESRange}

{\tt ESRange} is a parameter for the function {\tt DefES} controlling
the range of dummy indices. All dummy indices within \EinS\  can be of
two types:  "spatial" indices taking values {\tt
1,2,...,\$ESDimension-1}, and "space-time" indices taking values {\tt
0, 1,2,...,\$ESDimension-1}.  If {\tt ESRange -> \$ESDimension} is
specified as the third argument to {\tt DefES} all the dummy indices
specified in the second argument of {\tt DefES} are "space-time" ones.
If {\tt ESRange -> \$ESDimension-1} is specified (it is actually
assumed by default) all the dummy indices specified in the second
argument of {\tt DefES} are "spatial" ones.
If you have to specify indices of different types in a single
expressions nested calls of {\tt DefES} should be used.
The information on the type of each index is used, for example, by
the procedures {\tt ToComponents} and {\tt SplitTime} allowing
to convert an implicit sum into partially or fully explicit one.

See also: {\tt DefES}, {\tt \$ESDimension}, {\tt ToComponents},
{\tt SplitTime}

\ms\item[\begi]{\tt PD}
\index{PD}

\EinS\  has its own function for partial derivative. {\tt PD[x,y]}
represents partial derivative of {\tt x} with respect to {\tt y}. {\tt
PD[x,\{a,b,c,...\}]} represents multiple partial derivative of {\tt x}
with respect to {\tt a, b, c,...} The main reason to introduce one more
function for partial derivative in complement to \Mathematica's {\tt D}
was to make all objects dependent on all others by default.  Besides
that, the user can add addition rules to treat partial derivatives
within \EinS\  by specifying rules for {\tt PD} without danger to destroy
\Mathematica's engine by {\tt Unprotect}'ing {\tt D}.  Use {\tt
DeclareConstant[x]} or {\tt DeclareIndependent[x,n,{...}]} to make {\tt
x} independent of all or some variables. In current version of \EinS\
function {\tt PD} does not simplify derivatives of built-in
\Mathematica\ functions, such as {\tt Sin}.

See also: {\tt DeclareConstant}, {\tt DeclareIndependent}, {\tt DefRS}

\ms\item[\begi]{\tt DefRS}
\index{DefRS}

{\tt DefRS[coordinate\_name,time\_name]} "defines" coordinates by
generating simplification rules related with partial derivative {\tt PD}
as defined in \EinS, and producing output as well as \TeX\ formats.

See also: {\tt PD}, {\tt PrintES}, {\tt SaveTeX}

\ms\item[\begi]{\tt DeclareConstant}
\index{DeclareConstant}

{\tt DeclareConstant[object\_name,valence]} declares partial
derivatives {\tt PD} of specified object (as defined in \EinS) with
specified nonnegative valence with respect to any variable to be zero.

See also: {\tt PD}, {\tt DeclareIndependent}

\ms\item[\begi]{\tt DeclareIndependent}
\index{DeclareIndependent}

{\tt DeclareIndependent[object\_name,valence,list\_of\_variables]}
declares partial
derivatives {\tt PD} of specified object (as defined in \EinS) with
specified nonnegative valence with respect to specified variables to be
zero. Note that if the specified variable is the name of coordinates
defined previously by {\tt DefRS}, the rule generated by
{\tt DeclareIndependent}
concerns only "spatial" coordinate. If independence of coordinate time is
also desired it should be explicitly mentioned in the
{\tt list\_of\_variables}.

See also: {\tt PD}, {\tt DeclareConstant}

\ms\item[\begi]{\tt Delta}
\index{Delta}
\index{Mathematica 3.0}

{\tt Delta[i,j]} is the Kronecker symbol.  Its dimensionality can be
{\tt \$ESDimension} or {\tt \$ESDimension-1}.  {\tt Delta} is printed
as "$\delta$" under \Mathematica\ 3.0 notebook interface and as "{\tt
delta}" othewise. The \TeX\ format is always "{\tt $\backslash$delta}".

See also: {\tt \$ESDimension}, {\tt DropDelta}

\ms\item[\begi]{\tt LeviCivita}
\index{LeviCivita}
\index{Mathematica 3.0}

{\tt LeviCivita[i,j,...]} represents the fully antisymmetric
Levi-Civita symbol.  The number of arguments could be either {\tt
\$ESDimension} or {\tt \$ESDimension-1}. {\tt
LeviCivita[0,1,2,...,\$ESDimension]} as well as \hfil\break {\tt
LeviCivita[1,2,...,\$ESDimension]} are equal to {\tt 1}. Besides
anti-symmetry, and output ("$\varepsilon$" under \Mathematica\ 3.0
notebook interface and "{\tt eps}" othewise) and \TeX\ (always "{\tt
$\backslash$varepsilon}") formats, some simplification rules for
3-dimensional Levi-Civita symbols are built in \EinS.

See also: {\tt \$ESDimension}

\ms\item[\begi]{\tt c1}
\index{c1}

{\tt c1} represents $1/c$, $c$ representing normally speed of light.
This object should be used as small parameter in post-Newtonian
expansions. All procedures of \EinS\  are designed to be able to work
with series expansions with respect to any variable, but the object
{\tt c1} has some nice predefined output and \TeX\ formats.

\ei

\subsection{How to print expressions}
\label{print}

Having computed an expression you may want to look at it.
If you print at with \Mathematica's standard printing tools
(like, {\tt Print}) you will see internal \EinS's names for dummy indices,
which do not look "nice". \EinS\  has a built-in procedures enabling
one to print \EinS's expressions in "pretty" form.

\bi

\ms\item[\begi]{\tt PrintES}
\index{PrintES}
\index{Mathematica 3.0}

{\tt
PrintES[expression,\{list\_of\_nondummy\_indices\}]} {\tt PrintES}
prints the expression substituting internal names for dummy indices
with the strings from the list {\tt \$Indices} which are believed to be
"pretty" names for them. If the "pretty" name for a dummy index coincides
with actual name of a free (non-dummy) index entering in the same
expression, next member of {\tt \$Indices} is used instead.  Using the
second argument you can effectively exclude some of the "pretty" names
from {\tt \$Indices} which you prefer to use for other purposes (for
example, for free indices). Second argument is optional (default is
empty list).

Because of restricted printing capabilities of \Mathematica\ 2.x
(namely its inability to print greek characters), {\tt PrintES} prints
"space-time" dummy indices by adding simply "{\tt '}" to each index.
On the contrary, under \Mathematica\ 3.0 notebook interface, space-time
indices are always displayed as small greek letters. In this case the
\Mathematica\ strings with the symbols to be used to display space-time
dummy indices (presumably greak letters) are stored in the list {\tt
\$PGIndices}.  The same algorithm for selecting "pretty" names is used
for space-time dummy indices.

Note that two dummy indices in two different monomials may be printed
with the same "pretty" name while having different internal names.
Therefore, if you have printed with {\tt PrintES} two expressions
{\tt Ex1} and {\tt Ex2} and see visually that they are, say, equal,
\Mathematica\ does not necessarily automatically simplify {\tt Ex1-Ex2}
to {\tt 0}. Use {\tt ComputeES[Ex1-Ex2]} to simplify the difference.

\bs

\underbar{\sl Two notes for \Mathematica\ 3.0 notebook interface users:}
\index{Mathematica 3.0}

\begin{itemize}

\item
The new \Mathematica\ 3.0 enables one to use various {\tt FormatType}'s
to display output of your programs. Although any form could be used
with \EinS, our experience shows that {\tt OutputForm} is the most
appropriate to work with \EinS.  For the \Mathematica\ 3.0 notebook
front end one may want to change the default settings in {\tt Cell}
menu and set {\tt Default Output FormatType} to be {\tt OutputForm}.
In the same way you may want to change the default settings in {\tt Cell}
menu and set {\tt Default Input FormatType} to be {\tt InputForm} in order
to see \EinS\ input in its native form.

\item
Some versions of the \Mathematica\ 3.0 notebook front end contain a bug
which results sometimes in improper alignment of two dimensional
mathematical output if the expression contains greek characters or
other non-standard characters (say, superscripts and subscripts are
misaligned [shifted] with respect to the symbol carrying them). This
problem does not appear on print, but only on screen.  The only
work-around known to the author is to avoid naming object with
non-standard symbols. Let us hope that Wolfram Research, Inc. will fix
the bug as soon as possible.

\end{itemize}

\ms

See also: {\tt \$Indices}, {\tt \$PGIndices}, {\tt ComputeES}

\ms\item[\begi]{\tt \$Indices}
\index{\$Indices}

{\tt \$Indices} is a list of one-letter strings to be used as "pretty" names
for dummy indices when EinS outputs expressions using functions
{\tt PrintES}, {\tt SaveTeX} or {\tt EinS2TeX}.

See also: {\tt PrintES}, {\tt SaveTeX}

\ms\item[\begi]{\tt \$PGIndices}
\index{\$PGIndices}
\index{Mathematica 3.0}

{\tt \$PGIndices} is a list of one-letter strings to be used as
"pretty" names for space-time dummy indices when \EinS\ outputs
expressions using {\tt PrintES} under \Mathematica\ 3.0 notebook
interface.  By default, it contains all small greak letters in
alphabetical order.

See also: {\tt PrintES}

\ei

\subsection{How to compute and simplify expressions}
\label{compute}

\EinS\  has a variety of functions aimed at various simplifications of
expressions taking into considerations symmetries of user-defined as
built-in objects as well as opportunity to rename dummy indices.
Function {\tt ComputeES}, which performs standard maximal possible
simplification of an expression, is recommended for beginners. It
always returns the simplest form of the expression, but not necessarily
in minimal time. Other functions could be used by experienced users to
speed up the computations.

\bi

\ms\item[\begi]{\tt ComputeES}
\index{ComputeES}

{\tt ComputeES[expression]} performs standard maximal simplification possible
within \EinS. {\tt ComputeES[expression]} is equivalent to

\begin{verbatim}
UniqueES[
   ZeroBySymmetry[
      CollectES3[
         Normalize[
            ExpandAll[
               DropDelta[
                  Normalize[
                     ExpandSeries[
                        ExpandAll[expression]]]]]]]]]
\end{verbatim}

After standard simplification each dummy index in the resulting
expression has own unique name.

If {\tt \$ESDebug===True}, {\tt ComputeES} uses {\tt CheckDummies} to check if each
dummy index appear exactly twice in each term.

See also: {\tt UniqueES}, {\tt ZeroBySymmetry}, {\tt CollectES3},
{\tt Normalize}, {\tt DropDelta}, {\tt ExpandSeries}, {\tt CheckDummies},
{\tt \$ESDebug}

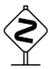
\ms\item[\expe]{\tt DropDelta}
\index{DropDelta}

{\tt DropDelta[expression]} is a procedure simplifying terms containing
Kronecker symbol {\tt Delta[i,j]} with a dummy index (or indices).

See also: {\tt Delta}

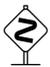
\ms\item[\expe]{\tt CollectES}
\index{CollectES}

{\tt CollectES[expression]} tries to match Einstein sums in the
expression renaming dummy indices.  This function is usually slower
than {\tt CollectES2}, but seems to be useful.  If {\tt
\$ESDebug===True} it prints numbers of equivalent terms.  {\tt
CollectES} should be invoked only after {\tt Normalize}. For historical
reasons, {\tt CollectES} does not account for possible antisymmetry of
tensors. Procedures {\tt CollectES}, {\tt CollectES2} and {\tt CollectES3}
represent 3 versions of the second part of simplification algorithm
implemented in \EinS, the first part being {\tt Normalize} or
{\tt MinimizeDummyIndices}, and the third one {\tt ZeroBySimmetry}.

See also: {\tt CollectES2}, {\tt CollectES3}, {\tt \$ESDebug}

\ms\item[\expe]{\tt CollectES2}
\index{CollectES2}

{\tt CollectES2[expression]} tries to match Einstein sums in the
expression renaming dummy indices. This function is much faster than
{\tt CollectES}. It uses internal Mathematica matching utility.  {\tt
CollectES2} does not account for possible antisymmetry of tensors.
If expression contains objects with antisymmetry as defined by
{\tt DefObject}, {\tt CollectES2} invokes automatically
{\tt CollectES3}. Procedures {\tt CollectES}, {\tt CollectES2} and {\tt
CollectES3} represent 3 versions of the second part of simplification
algorithm implemented in \EinS, the first part being {\tt Normalize} or
{\tt MinimizeDummyIndices}, and the third one {\tt ZeroBySymmetry}.

See also: {\tt CollectES3}, {\tt CollectES}

\ms\item[\expe]{\tt CollectES3}
\index{CollectES3}

{\tt CollectES3[expression]} tries to match Einstein sums in the
expression renaming dummy indices. {\tt CollectES3} is equivalent to
{\tt CollectES2}, but accounts for antisymmetry of tensors.
If expression does not contain objects with antisymmetry as defined by
{\tt DefObject}, {\tt CollectES3} invokes automatically
{\tt CollectES2}. Procedures {\tt CollectES}, {\tt CollectES2} and {\tt
CollectES3} represent 3 versions of the second part of simplification
algorithm implemented in \EinS, the first part being {\tt Normalize} or
{\tt MinimizeDummyIndices}, and the third one {\tt ZeroBySimmetry}.

See also: {\tt CollectES2}, {\tt CollectES}

\ms\item[\expe]{\tt UniqueES}
\index{UniqueES}

{\tt UniqueES[expression]} renames all dummy indices of the expression
to have unique names. It is necessary for making substitutions of
{\tt Normalize}'d expressions into other expressions.

See also: {\tt Normalize}

\ms\item[\expe]{\tt Normalize}
\index{Normalize}

{\tt Normalize[expression]} renames all dummy indices in the expression
so that number of different dummy indices is minimal for each Einstein
sum, and factors all terms independent of dummy indices out of the
Einstein sum. Dummy indices will be numbered in the order they appear
in the expression itself (which is ordered lexicographically by
\Mathematica\ itself). {\tt Normalize[expression]} is NOT equivalent to
{\tt MinimizeDummyIndices[FactorConstants[expression]]}.
Procedures {\tt Normalize} and {\tt MinimizeDummyIndices} represent 2
versions of the first part of simplification algorithm implemented in
\EinS, the second part being {\tt CollectES}, {\tt CollectES2} or {\tt
CollectES3}, and the third one {\tt ZeroBySimmetry}. Many terms
involving implicit summation are combined automatically by \Mathematica
itself after applying {\tt Normalize} or {\tt MinimizeDummyIndices}.

See also: {\tt UniqueES}, {\tt MinimizeDummyIndices}, {\tt FactorConstants}

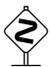
\ms\item[\expe]{\tt MinimizeDummyIndices}
\index{MinimizeDummyIndices}

{\tt MinimizeDummyIndices[expression]} renames all dummy indices in the
expression so that number of different dummy indices is minimal for
each Einstein sum. Dummy indices will be numbered in the order they appear
in the lists of dummy indices (second arguments of {\tt ES}).
Procedures {\tt Normalize} and {\tt MinimizeDummyIndices} represent 2
versions of the first part of simplification algorithm implemented in
\EinS, the second part being {\tt CollectES}, {\tt CollectES2} or {\tt
CollectES3}, and the third one {\tt ZeroBySimmetry}. Many terms
involving implicit summation are combined automatically by \Mathematica
itself after applying {\tt Normalize} or {\tt MinimizeDummyIndices}.
Note that {\tt MinimizeDummyIndices} renames dummy indices in another
order a compared with {\tt Normalize} which can be of advantage or
disadvantage depending on the problem under study. {\tt Normalize} is used
in the standard simplification function {\tt ComputeES}.

See also: {\tt UniqueES}, {\tt Normalize}, {\tt ES}

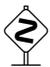
\ms\item[\expe]{\tt FactorConstants}
\index{FactorConstants}

{\tt FactorConstants[expression]} factors all terms independent of
dummy indices out of each Einstein sum in the expression.

See also: {\tt Normalize}

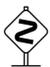
\ms\item[\expe]{\tt ZeroBySymmetry}
\index{ZeroBySymmetry}

{\tt ZeroBySymmetry[expression]} tests each term of the expression for
being equal to zero due to be a product of symmetric and antisymmetric
factors for a pair of dummy indices.  Procedure {\tt ZeroBySymmetry}
represent the third part of simplification algorithm implemented in
\EinS, the first part being {\tt Normalize} or {\tt
MinimizeDummyIndices}, and the third one {\tt CollectES}, {\tt
CollectES2} or {\tt CollectES3}. {\tt ZeroBySymmetry} may take
relative long running time for large number of dummy indices in terms
involving many objects with defined antisymmetry.

See also: {\tt ComputeES}

%

\ei

\subsection{How to convert implicit sums into partially and fully
explicit form}
\label{converttoexplicit}

\EinS\  has two functions which allow to convert implicit summations into
partially ("3+1" split in case of general relativity) or fully explicit
form. The latter is a conversion from indicial tensor notations into
components notations which is useful for numerical computing as well as
for checking the results.

\bi

\ms\item[\begi]{\tt SplitTime}
\index{SplitTime}

{\tt SplitTime[expression]} represents any Einstein sum involving
"space-time" indices into sum of corresponding terms with zero values of
indices and the sum involving "spatial" indices.

See also: {\tt ToComponents}, {\tt ESRange}

\ms\item[\begi]{\tt ToComponents}
\index{ToComponents}

{\tt ToComponents[expression]} converts implicit Einstein summation
into explicit form. In the explicit form all indices are integer
numbers.  All dummy indices are considered to run either from {\tt 0}
(for "space-time" indices) or from {\tt 1} (for "spatial" indices) to
{\tt \$ESDimension-1}.

See also: {\tt SplitTime}, {\tt \$ESDimension}, {\tt ESRange}

\ei

\subsection{How to export expression into \TeX}
\label{exporttoTeX}

\EinS\  has a built-in subsystem allowing to export \EinS's expressions
into \TeX\ or \LaTeX\ with automatic line breaking of formulas with
specified number of monomials per line.

\bi

\ms\item[\begi]{\tt SaveTeX}
\index{SaveTeX}

{\tt SaveTeX[file\_name,expression,{list\_of\_nondummy\_indices}]}
outputs a string returned by
\hfil\break
{\tt EinS2TeX[expression,{list\_of\_nondummy\_indices}]}
\hfil\break
to the specified file. The meaning of the third argument is the same as
for the procedure {\tt PrintES}. The third argument is optional and
empty by default.

See also: {\tt EinS2TeX}, {\tt \$TeXDialect}

\ms\item[\begi]{\tt EinS2TeX}
\index{EinS2TeX}

{\tt EinS2TeX[expression,{list\_of\_nondummy\_indices}]} returns a
string containing TeX form of the expression according to switches {\tt
\$TeXDialect}, {\tt \$TermsPerLine}, and the list of "pretty" names for
"spatial" dummy indices {\tt \$Indices} and "space-time" dummy indices
{\tt \$GIndices}. The meaning of the second argument is the same as
for the procedure {\tt PrintES}.  The second argument is optional and
empty by default.

See also: {\tt SaveTeX},{\tt \$TeXDialect}, {\tt \$TermsPerLine}
{\tt \$Indices}, {\tt \$GIndices}

\ms\item[\begi]{\tt \$GIndices}
\index{\$GIndices}

\nobreak

{\tt \$GIndices} is a list of strings to be used as "pretty" names for
"space-time" dummy indices when \EinS\  outputs expressions using functions
{\tt SaveTeX} or {\tt EinS2TeX}.
By default, it contains \TeX\ forms of small greek letters in alphabet order.

See also: {\tt SaveTeX}, {\tt EinS2TeX}

\ms\item[\expe]{\tt EinS2TeXString}
\index{EinS2TeXString}

{\tt EinS2TeXString[expression]:=string}

A set of rules with the head {\tt EinS2TeXString} is used when converting
\EinS's expressions into their \TeX\ representation.  Besides a
number of predefined rules, the procedure {\tt DefObject} automatically
adds rules for user-defined objects. User can add his own rules
manually which you cannot for some reason define via {\tt
DefObject}. {\tt EinS2TeXString} should return a \Mathematica's string.

See also: {\tt SaveTeX}, {\tt EinS2TeX}, {\tt DefObject}

\ms\item[\begi]{\tt \$TeXDialect}
\index{\$TeXDialect}

{\tt \$TeXDialect} controls whether {\tt EinS2TeX} and {\tt SaveTeX}
output EinS expressions with alignment commands
in plain \TeX\ ({\tt \$TeXDialect===\$PlainTeX}), \LaTeX\
({\tt \$TeXDialect===\$LaTeX})
or without alignment
({\tt \$TeXDialect===None}). In the latter case output is not split
into lines and no alignment is performed. Default is {\tt \$LaTeX}.

\ms\item[\begi]{\tt \$PlainTeX}
\index{\$PlainTeX}

{\tt \$PlainTeX} is a possible value for {\tt \$TeXDialect}. If
{\tt \$TeXDialect===\$PlainTeX}
{\tt EinS2TeX} and {\tt SaveTeX} output \EinS's expressions into
plain \TeX\ form.

\ms\item[\begi]{\tt \$LaTeX}
\index{\$LaTeX}

{\tt \$LaTeX} is possible value for {\tt \$TeXDialect}.
If {\tt \$TeXDialect===\$LaTeX}
{\tt EinS2TeX} and {\tt SaveTeX} output \EinS's expressions into
\LaTeX\ form.

\ms\item[\begi]{\tt \$TermsPerLine}
\index{\$TermsPerLine}

{\tt \$TermsPerLine} is an integer number of terms to be printed per
line when formatting multiline \TeX\ or \LaTeX\ output using functions
{\tt EinS2TeX} or {\tt SaveTeX}.

\ei

\subsection{How to debug programs}
\label{debug}

\EinS\  has limited debugging capabilities. It includes one switch and
one procedure.

\bi

\ms\item[\begi]{\tt \$ESDebug}
\index{\$ESDebug}

{\tt \$ESDebug} is a switch turning on and off printing some additional
information from several functions defined in \EinS. Many of messages
are of form

\begin{verbatim}
FunctionName:: ??? terms processed
\end{verbatim}

\noindent
and are printed each 100 terms. Here {\tt FunctionName} stands for the
name of the \EinS's function which prints current message, and {\tt
???} for actual number of terms. These messages are useful for checking
what \EinS\  is doing now.

Beside that, if {\tt \$ESDebug===True} {\tt ComputeES} and {\tt DefES}
automatically call {\tt CheckDummies} to check the validity of the
usage of dummy indices.

See also: {\tt CheckDummies}

\ms\item[\begi]{\tt CheckDummies}
\index{CheckDummies}

{\tt CheckDummies[expression]} checks that in each term of expression
each dummy index appears exactly two times and print a message
otherwise.  If {\tt \$ESDebug==True}, {\tt CheckDummies} is invoked
automatically by {\tt ComputeES} and {\tt DefES}.

See also: {\tt \$ESDebug}, {\tt DefES}, {\tt ComputeES}

\ei

\bs

Many of the functions generate errors or warning messages if something
goes wrong. If you got an error message you have to check, for example,
that you are doing a meaningful operation. If you believe that it is
meaningful and cannot find an error in your code, you may want to
check if each pair of dummy indices during you calculations
does have a unique name. In many cases \EinS\  tries to generate
unique names automatically and it is always possible to write the program
in such a way that \EinS\  does this itself. However, if you think
\EinS\  should do this in your case, but does not do this,
you may want to make this manually by using function {\tt UniqueES}.

\subsection{Some internal commands}
\label{internal}

This section describes some of the internal commands of \EinS\  which we
decided to make accessible to the users. These functions allow to write
subroutines extending functionality of the package. With except for
{\tt \$MaxDummy}, these functions should be used only by experienced
users.

\bi

\ms\item[]{\tt \$ESVersion}
\index{\$ESVersion}

A string representing EinS logo and version.

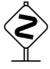
\ms\item[\expe]{\tt ES}
\index{ES}

{\tt ES} serves as a token folding each Einstein sum.
\hfil\break
{\tt ES[expression,list\_of\_dummy\_indices} is internal representation
of Einstein sum of expression with respect to the dummy indices from
the list. Normally {\tt ES} should not be used by the user. Dummy indices
are coded as {\tt \$DummyHead[number]} for "space-time" dummy indices, and
{\tt \$SDummyHead[number]} for "spatial" dummy indices, where
{\tt number} is number of the pair of dummy indices. This {\tt number}
is usually unique for each pair of dummy indices processed by \EinS,
but can be changed by functions {\tt Normalize}, {\tt UniqueES} and
{\tt MinimizeDummyIndices}

See also: {\tt Normalize}, {\tt UniqueES}, {\tt MinimizeDummyIndices}

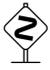
\ms\item[\expe]{\tt STDummyQ}
\index{STDummyQ}

{\tt STDummyQ[expression]} returns {\tt True} if the expression is a
"space-time" dummy index and {\tt False} otherwise.

See also: {\tt DummyQ}, {\tt SDummyQ}

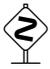
\ms\item[\expe]{\tt SDummyQ}
\index{SDummyQ}

{\tt SDummyQ[expression]} returns {\tt True} if the expression is a "spatial"
dummy index and {\tt False} otherwise.

See also: {\tt DummyQ}, {\tt STDummyQ}

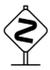
\ms\item[\expe]{\tt DummyQ}
\index{DummyQ}

{\tt DummyQ[expression]} returns {\tt True} if the expression is
a dummy index and {\tt False} otherwise.

See also: {\tt SDummyQ}, {\tt STDummyQ}

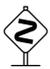
\ms\item[\expe]{\tt \$\$Alt}
\index{\$\$Alt}

A symbol folding indices for which an expression is alternating.

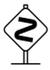
\ms\item[\expe]{\tt \$\$Sym}
\index{\$\$Sym}

A symbol folding indices for which an expression is symmetric.

\ms\item[\begi]{\tt \$MaxDummy}
\index{\$MaxDummy}

Maximal number of dummy indices to be handle by EinS.
EinS performance does not depend on {\tt \$MaxDummy}.
By default,  {\tt \$MaxDummy=1000}.

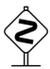
\ms\item[\expe]{\tt \$DummyHead}
\index{\$DummyHead}

A string used by EinS as Head for internal representation of "space-time"
dummy indices. Initially "a\$\$" (a \Mathematica\ string).

See also: {\tt \$SDummyHead}

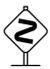
\ms\item[\expe]{\tt \$SDummyHead}
\index{\$SDummyHead}

A string used by EinS as Head for internal representation of "spatial"
dummy indices. Initially "i\$\$" (a \Mathematica\ string).

See also: {\tt \$DummyHead}

\ei

\newpage

\section{Future prospects}
\label{future}

As we mentioned above, \EinS\  resulted from our own needs. It was our
desire to have a relatively small, simple and easy-to-tune package
allowing to perform calculations aroused from various scientific
problems the author investigated. Therefore, it is generally not our
intention to improve package permanently. On the contrary, the
further development of the package strongly depend on the scientific
problems which the author will study.
Nevertheless, we list the improvements of \EinS\  which are to be expected
in the future:

\bi

\item
Refining simplification algorithm in two major directions:

\be

\item[-]
automated splitting of dummy indices into subgroups which cannot
intersect a priori when performing pattern matching;

\item[-] handling more complicated symmetry properties including
linear (and possibly non-linear) identities (see, e.g.,
\cite{rodtar:87,ilkry:91,ilkry:94}).

\ee

\item

 Writing a special package on the basis on \EinS\  for calculation of a
standard set of quantities on the basis of metric (Christoffel symbols,
curvature tensor, covariant derivative, etc.). It is clear that
implementing such a package is rather trivial task having \EinS.  At
present time, many potential pieces of such a package is distributed
over various applications of \EinS. However, we probably will not make
\EinS\  to distinguish automatically covariant and contravariant indices.

\ei

Besides that, some minor changes could be made on request of users,
if such changes are in agreement with the author's vision of the \EinS's
functionality. You are encouraged to report bugs (which certainly should
exist in the package) as well as suggestions to the address quoted on
the title page of this document.

\section{Acknowledgements}

The version 2.6 of \EinS\  described in this document has been finished
while the author was Humboldt Research Fellow at the Lohrmann
Observatorium, Dresden, Germany. Financial support of the Alexander
von Humboldt Foundation as well as hospitality of Prof. Michael
Soffel and the stuff of the Lohrmann Observatorium are gratefully
acknowledged.

\newpage

\addcontentsline{toc}{section}{Index}

\begin{theindex}

  \item {\tt \$DummyHead}, 22
  \item {\tt \$ESDebug}, 20
  \item {\tt \$ESDimension}, 9
  \item {\tt \$ESVersion}, 21
  \item {\tt \$GIndices}, 19
  \item {\tt \$Indices}, 14
  \item {\tt \$LaTeX}, 19
  \item {\tt \$MaxDummy}, 22
  \item {\tt \$PGIndices}, 14
  \item {\tt \$PlainTeX}, 19
  \item {\tt \$SDummyHead}, 22
  \item {\tt \$TeXDialect}, 19
  \item {\tt \$TermsPerLine}, 19
  \item {\tt \$\$Alt}, 21
  \item {\tt \$\$Sym}, 21

  \indexspace

  \item {\tt c1}, 13
  \item {\tt CheckDummies}, 20
  \item {\tt CollectES}, 15
  \item {\tt CollectES2}, 16
  \item {\tt CollectES3}, 16
  \item {\tt ComputeES}, 15

  \indexspace

  \item {\tt DeclareConstant}, 12
  \item {\tt DeclareIndependent}, 12
  \item {\tt DefES}, 11
  \item {\tt DefObject}, 9
  \item {\tt DefRS}, 12
  \item {\tt Delta}, 12
  \item {\tt DropDelta}, 15
  \item {\tt DummyQ}, 21

  \indexspace

  \item {\tt EinS2TeX}, 18
  \item {\tt EinS2TeXString}, 19
  \item {\tt ES}, 21
  \item {\tt ESIndicesLow}, 11
  \item {\tt ESRange}, 11
  \item {\tt ExpandSeries}, 9

  \indexspace

  \item {\tt FactorConstants}, 17

  \indexspace

  \item {\tt KSubsets}, 9

  \indexspace

  \item {\tt LeviCivita}, 13
  \item Loading \EinS, 7

  \indexspace

  \item \Mathematica\ 3.0, 7, 12--14
  \item {\tt MinimizeDummyIndices}, 17

  \indexspace

  \item {\tt Normalize}, 16

  \indexspace

  \item {\tt PD}, 11
  \item {\tt PrintES}, 13

  \indexspace

  \item {\tt SaveDefinition}, 8
  \item {\tt SaveExpression}, 8
  \item {\tt SaveTeX}, 18
  \item {\tt SDummyQ}, 21
  \item Series within \EinS, 8, 9
  \item {\tt SeriesLength}, 8
  \item {\tt SplitTime}, 18
  \item {\tt STDummyQ}, 21

  \indexspace

  \item {\tt ToComponents}, 18

  \indexspace

  \item {\tt UniqueES}, 16

  \indexspace

  \item {\tt ZeroBySymmetry}, 17

\end{theindex}

\end{document}